\documentclass[aps,twocolumn,prl,showpacs,preprintnumbers,amsmath,amssymb,superscriptaddress]{revtex4}
\usepackage{mathptm}  
\usepackage{dcolumn}                    
\usepackage{bm}                        
\usepackage{graphicx}
\usepackage{times}
\usepackage{graphicx}
\begin{document}
\newcommand {\be}{\begin{equation}}
\newcommand {\ee}{\end{equation}}
\newcommand {\bea}{\begin{eqnarray}}
\newcommand {\eea}{\end{eqnarray}}
\newcommand {\nn}{\nonumber}
\newcommand {\bb}{\bibitem}
\small

\title{Small T$_{1}^{-1}$ coherence peak near T$_{c}$ in unconventional BCS
superconductors}

\author{David Parker}
\affiliation{Department of Physics and Astronomy, University of Southern
California, Los Angeles, CA 90089-0484 USA}
\affiliation{Max Planck Institute for the Physics of Complex Systems,
N\"{o}thnitzer Str. 38, D-01187 Dresden, Germany}

\author{Stephan Haas}

\affiliation{Department of Physics and Astronomy, University of Southern
California, Los Angeles, CA 90089-0484 USA}

\date{\today}

\begin{abstract}
It is usually believed that a coherence peak just below T$_{c}$ 
in the nuclear spin lattice relaxation rate T$_{1}^{-1}$ in superconducting
materials is a signature of conventional s-wave pairing.  In this paper we
demonstrate that {\bf any} unconventional superconductor 
obeying BCS pure-case 
weak-coupling theory should show a small T$_{1}^{-1}$ coherence peak 
near T$_{c}$, generally with a height between 3 and 15 percent greater than the
normal state T$_{1}^{-1}$ at T$_{c}$.  It is largely due to impurity and magnetic effects that this peak has not commonly been observed.  
\end{abstract}
\maketitle

It is well known \cite{schrieffer} that conventional superconductors obeying BCS
weak-coupling theory generally show a large coherence peak in the nuclear spin
lattice relaxation rate T$_{1}^{-1}$
below T$_{c}$.  This is a direct consequence of the large
quasi-particle density-of-states found for $E \geq 1.0$ at the gap edge.
It is commonly believed, and experimentally largely true, that unconventional superconductors show no
coherence peak near T$_{c}$.  However, a simple quantitative argument shows 
that if the BCS weak-coupling pure case theory applies, a small coherence peak below
T$_{c}$ must exist.

Consider the BCS weak-coupling equation for the nuclear spin lattice relaxation
rate T$_{1}^{-1}$ for unconventional superconductors:
\bea
(T_{1} T)^{-1}/(T_{1} T)^{-1}_{|T=T_{c}} =
\int_{0}^{\infty} dE\,\, N^{2}(E)
\mathrm{sech}^{2}(E/(2T))/2T
\eea
The $\mathrm{sech}^{2}(E/(2T))$ acts as an attenuation factor and dominates
the low-temperature T$_{1}^{-1}$, yielding exponentially activated
behavior for s-wave superconductivity and power-law behavior for 
unconventional superconductivity.  However, as T $\rightarrow T_{c}$ all
of the structure in N(E) (i.e. DOS different from unity) is shifted to lower
energies, since $N(E)= N(E/\Delta)$ and $\Delta \rightarrow 0$.  
See Figure 1 for a depiction of this behavior for the 3-d $^3$-He A-phase 
order parameter
$\Delta({\bf k})=\Delta \sin\theta$.  All of the structure in N(E)
\begin{figure}[h!]
\includegraphics[width=6.5cm]{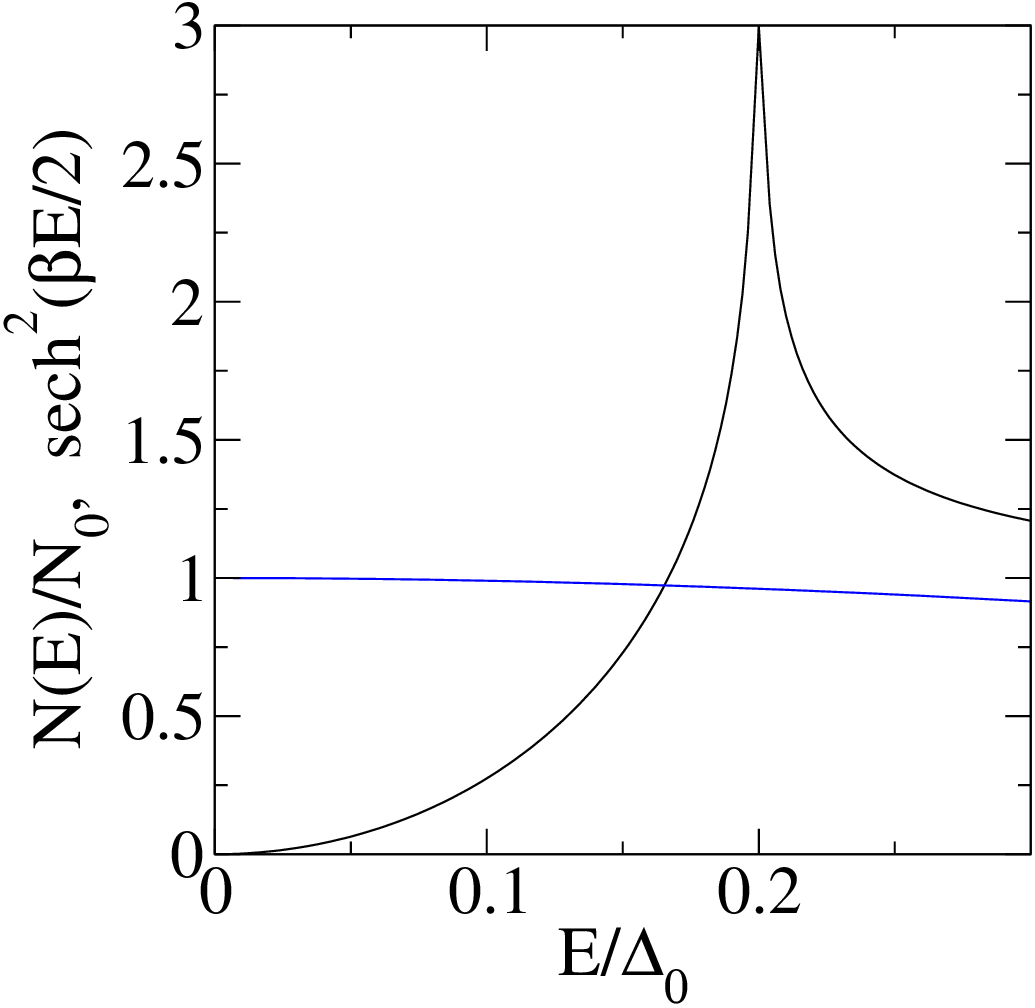}
\caption{Diagram depicting attenuation factor and quasi-particle
density-of-states for $T \simeq 0.98 T_{c} \simeq 0.5 \Delta_{0}, \Delta=
0.2\Delta_{0}$, 3-d $^3$-He A-phase order parameter.}
\end{figure}
falls in the region where $E/2T \ll 1$, so that for this structure the
exponential damping factor becomes essentially irrelevant.  Define 
$F(E)= N(E)-1$.  Note that by the density-of-states sum rule, $\int_{0}^{\infty}
F(E) dE = 0$.  Now substituting in for N(E) we find
\bea
(T_{1} T)^{-1}/(T_{1} T)^{-1}_{|T=T_{c}} =  \nonumber \\
\int_{0}^{\infty} dE (1+2F(E)+F^
{2}(E)) \mathrm{sech}^{2}(E/(2T))/2T
\eea
The first term trivially yields 1.  The second
term, $\int_{0}^{\infty} dE\,\, 2F(E) \mathrm{sech}^{2}(E/(2T))/2T$ can be 
evaluated 
by noticing that as $\Delta \rightarrow 0$, F(E) is only significantly
different from 0 in regions where the argument of the $\mathrm{sech}^{2}$ is
small, so that to an excellent approximation near T$_{c}$ this integral is equal to
$\int_{0}^{\infty} F(E)\,\, dE$ = 0.  The final term, $\int_{0}^{\infty} dE\,\, F^{2}(E) \mathrm{sech}^{2}(E/(2T))/2T$, is positive, and so in the immediate neighborhood of T$_{c}$, $(T_{1} T)^{-1}/(T_{1} T)^{-1}_{|T=T_{c}} > 1$,
implying the existence of a peak. 
It is this redistribution of N(E) away
from an energy-constant ($=N_{0}$) density-of-states, represented
by $F^{2}(E)$, that is responsible for the
peak in $T_{1}^{-1}$ near T$_{c}$.  The larger this effect, the
larger the peak.

This redistribution is intimately tied in with the nodal structure of 
$\Delta({\bf k})$.  This can be seen directly from 
the BCS expression for the
density-of-states $N(E/\Delta) \equiv N(x) = \mathrm{Re}\langle\frac{x}{\sqrt{x^2-f^2}}
\rangle$, where $\langle \ldots \rangle$ denotes an average over the
Fermi surface and f contains the angular dependence of the order parameter. 
(i.e., $\Delta({\bf k}) = \Delta_{0}f({\bf k})$).  The contribution of the 
nodes is most easily parametrized by $<f^{2}>$, with larger values
indicating less nodal order parameters.  For example, an s-wave order parameter
has $<f^{2}>$ = 1, while a 2d d-wave order parameter (containing line nodes) has $<f^{2}>$ =0.5.  Gap
functions $f$ with larger $<f^{2}>$, indicating effectively small or absent nodes, have a comparatively smaller region of phase space contributing to the integral, for $x< 1$.  These gap functions 
will therefore show depleted low-energy density-of-states, and by the sum rule
must have enhanced spectral weight in the peak at $E=\Delta$. Both effects will tend to enhance the T$_{1}^{-1}$ peak just below T$_{c}$.

These behaviors are illustrated in Figure 2, which depicts 
densities-of-states and $T_{1}^{-1}$ for a series of 3-d order parameters
$\Delta({\bf k})=1-\cos^{n}(\theta)$, with $\theta$ the polar angle.  As n
increases, the low-energy DOS is depleted and the coherence-peak DOS enhanced,
with a concomitant increase in the $T_{1}^{-1}$ peak near T$_{c}$.  For these
cases, $<f^{2}>$ increases monotonically from $\frac{8}{15} \simeq 0.533$ for n=2 to 0.866 for n=10.
\begin{figure}[h!]
\includegraphics[width=8cm]{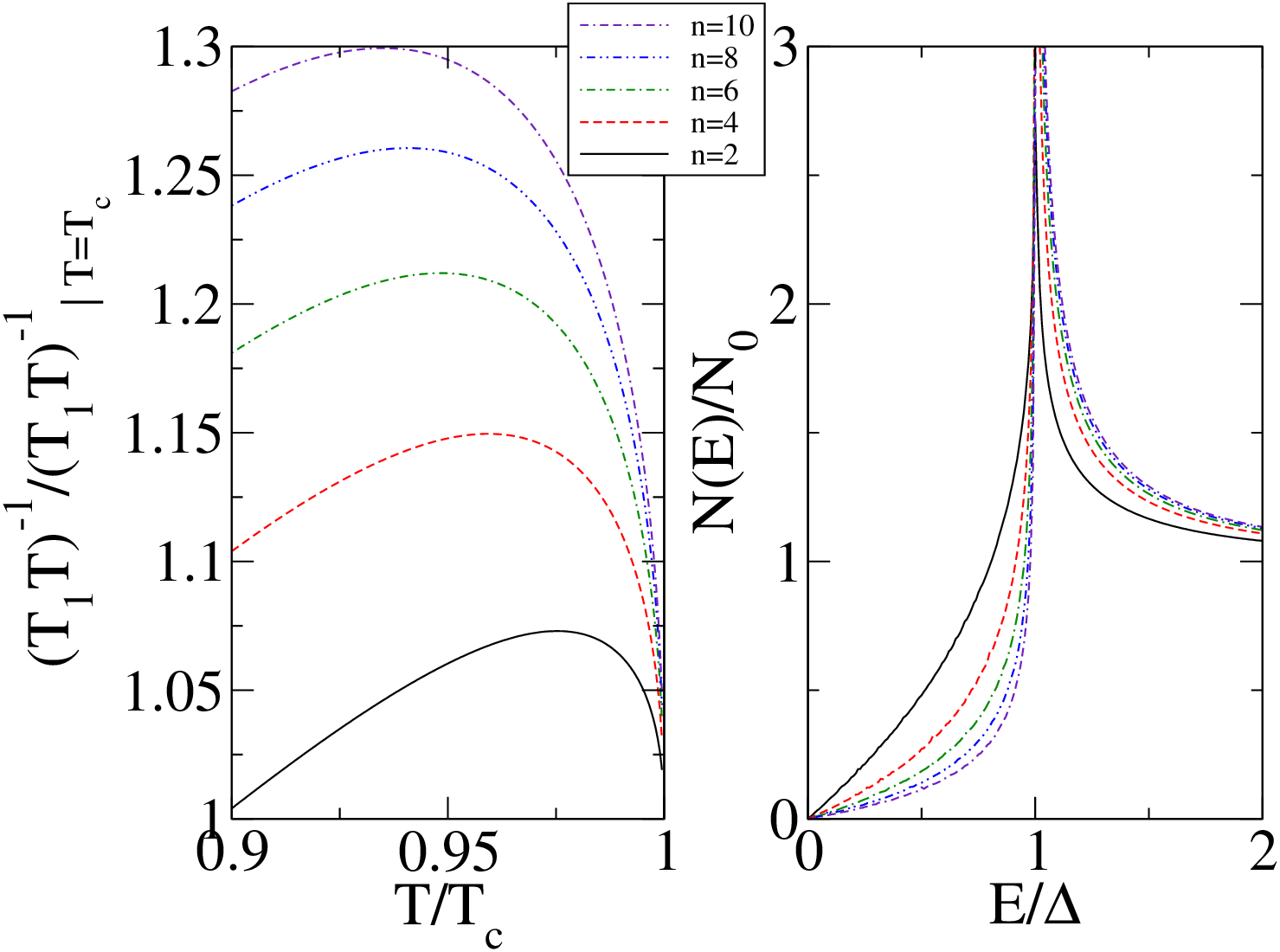}
\caption{The densities-of-states and T$_{1}^{-1}$ for several 3D 
order parameters $\Delta({\bf k})=1-\cos^{n}(\theta)$ are shown.}
\end{figure}

{\it Results.-} In Figure 3 are depicted the coherence peaks near T$_{c}$ for several
unconventional order parameters: d$_{x^{2}-y^{2}}$-wave, the $^{3}$-He A-phase
order parameter (for which $\Delta({\bf k})=\Delta_{0}\sin\theta$), and the p-wave 3-dimensional order parameter $\Delta({\bf k})=
\Delta_{0}\cos\theta$, as well as the quasiparticle density-of-states
\begin{figure}[h!]
\includegraphics[width=8cm]{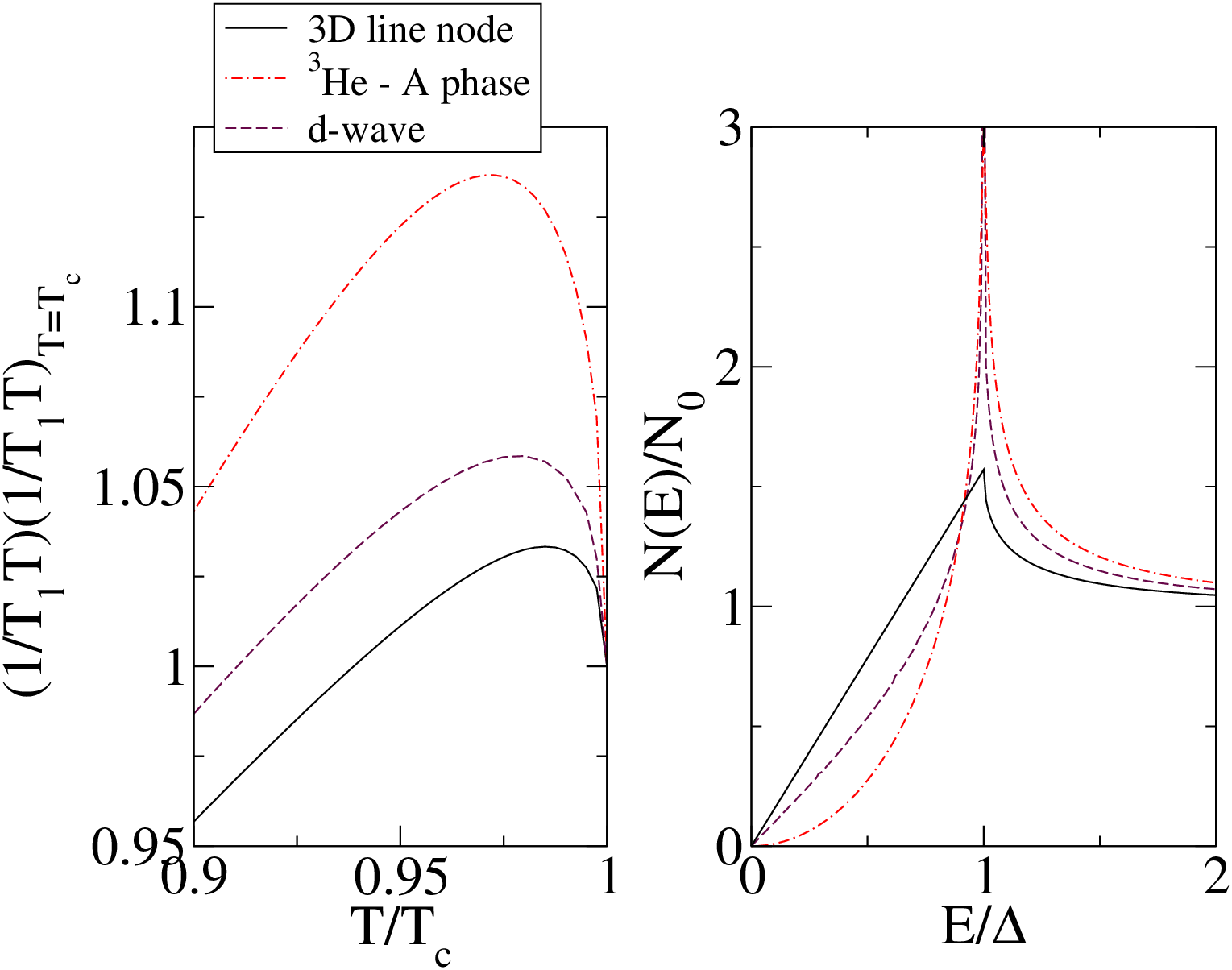}
\caption{The predicted 
nuclear spin lattice relaxation rates (T$_{1}T)^{-1}$ near T$_{c}$ and quasiparticle
densities-of-states
are shown for three order parameters: 2D d-wave, the A phase of $^3He$, and a 
3D line-node model.}
\end{figure}
for these order parameters.  Note that a small peak 
just below T$_{c}$ is evident even for the last order parameter, 
whose quasiparticle density-of-states shows no divergence at $E=\Delta$.  The
analysis of the preceding paragraph demonstrates that it is largely
the second moment of the DOS around an energy-constant DOS ($\simeq N_{0}$)
that produces the small peak just below T$_{c}$.  
While a large or divergent DOS at $E=\Delta$ clearly enhances 
the coherence peak near T$_{c}$, it is not necessary for the formation
of a peak.

For the three cases described above, it is possible to derive an analytic
expression for T$_{1}^{-1}$ just below T$_{c}$ and compare with the numerical
results.  The quasiparticle density-of-states for each of the three order
parameters can be computed analytically, and one finds the following well known
results \cite{maki,hirschfeld,sigrist}, where $x=E/\Delta$ and K is the elliptical function:
\bea
\mathrm{d-wave:\,} N(x) & = & \frac{2}{\pi}K(\frac{1}{x}), x<1; \\
&=& \frac{2}{\pi}x K(x), x>1 \\
\mathrm{^{3}He\,\, A-phase:\,} N(x) &=& \frac{x}{2}\log(|\frac{1+x}{1-x}|) \\
\mathrm{3d-line node:\,} N(x) &=& \frac{\pi}{2}x, x \leq 1; \\
&=& x \sin^{-1}(\frac{1}{x}), x \geq 1.
\eea

Now, to work out an analytic form for the peak in
T$_{1}^{-1}$ just below T$_{c}$, we must 
compute $\int_{0}^{\infty} dE F^{2}(E) \mathrm{sech}^{2}(E/(2T))/2T$, where
$F(E)=N(E)-1$.  For T sufficiently near T$_{c}$, F(E) only varies from
zero in a region where $\mathrm{sech}^{2}(E/2T)$ is essentially unity, as described
at the beginning of this paper.  Making the substitution $E=\Delta x$, and
taking $\mathrm{sech}^{2}(E/2T)$ as 1, 
we find for the above integrals, where $\gamma$ is the
Catalan constant$= 0.915\ldots$
\bea
\mathrm{d-wave:} & \simeq & \,\,\,0.4743\Delta/2T \\
\mathrm{^{3}-He A-phase:} \,\,\, \frac{\pi^{2}}{12}\Delta/2T 
&\simeq & 0.822\Delta/2T \\
\mathrm{3D\,\,line\,\,node:} \,\,\,\ (\frac{2}{3}\gamma - \frac{1}{3}) \Delta/2T &\simeq&
0.2773\Delta/2T
\eea
In other words, very near T$_{c}$ ($T > 0.995 T_{c}$) we can express the 
ratio $(T_{1} T)^{-1}/(T_{1} T)^{-1}_{|T=T_{c}}$ as simply 1+$\alpha \Delta(T)/T$, where $\alpha$
is an order parameter-dependent constant, and
this expression yields reasonably good agreement with the numerical results.  
In order to better model the behavior near T$_c$ we have calculated analytically the next order term and found excellent agreement, 
as indicated in the plot below.  Below 0.98 T$_{c}$ this approximation becomes less accurate.
\begin{figure}[h!]
  \includegraphics[width=6.5cm]{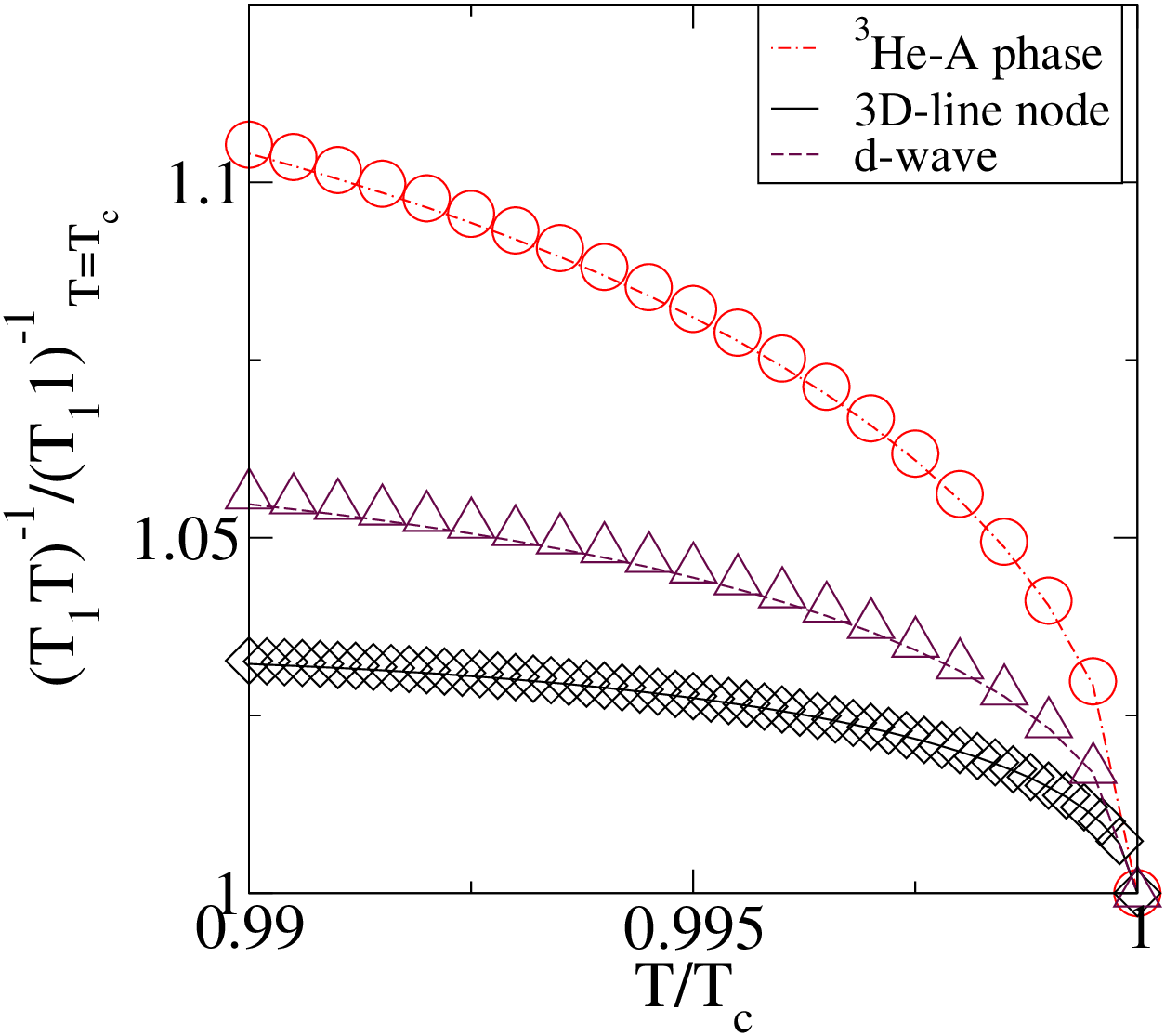}
\caption{The analytic and numerically computed T$_{1}^{-1}$ for the three
order parameters previously discussed are shown. Symbols indicate the analytic
result and the lines the numerical result.}
\end{figure}

{\it Discussion.-} The foregoing analysis shows that BCS weak-coupling
pure-case unconventional superconductors should exhibit a small T$^{-1}_{1}$
coherence peak just below T$_{c}$.  Yet a literature survey on this topic 
\cite{lee,kotegawa,tien,ishida,maclaughlin,zheng,matsuda,tien2,kawasaki,
fujimoto,curro,ishida2,kanoda,desoto,kotegawa2,kohori,kawasaki2,kato,
maclaughlin2,iwamoto,sakai,mito} has uncovered just two materials  - CePt$_{3}$Si and
(TMTSF)$_{2}$PF$_{6}$-which 
show such a peak.  The question therefore arises as to why such peaks are
not commonly observed.

To address this question, we have conducted an analysis of the effect of
resonant impurity scattering upon this T$_{1}^{-1}$ peak, for a 
two dimensional d-wave order parameter.  It is well known that such
impurity scattering truncates the DOS peak at $E=\Delta$ and in addition can
generate substantial low-energy density-of-states.  Both of these effects
would tend to reduce the size of the coherence peak.  It turns out that 
for these reasons the appearance of this peak is extraordinarily sensitive
to impurity scattering.  Depicted below are four T$_{1}^{-1}$ curves for
d-wave superconductivity: zero impurity scattering, and three cases of small
impurity scattering: $\Gamma/\Delta_{00} = 0.01$, $0.02$ and $0.03$.  Within the unitary
limit this last concentration is roughly 7 percent of the critical impurity
\begin{figure}[h!]
\includegraphics[width=6.5cm]{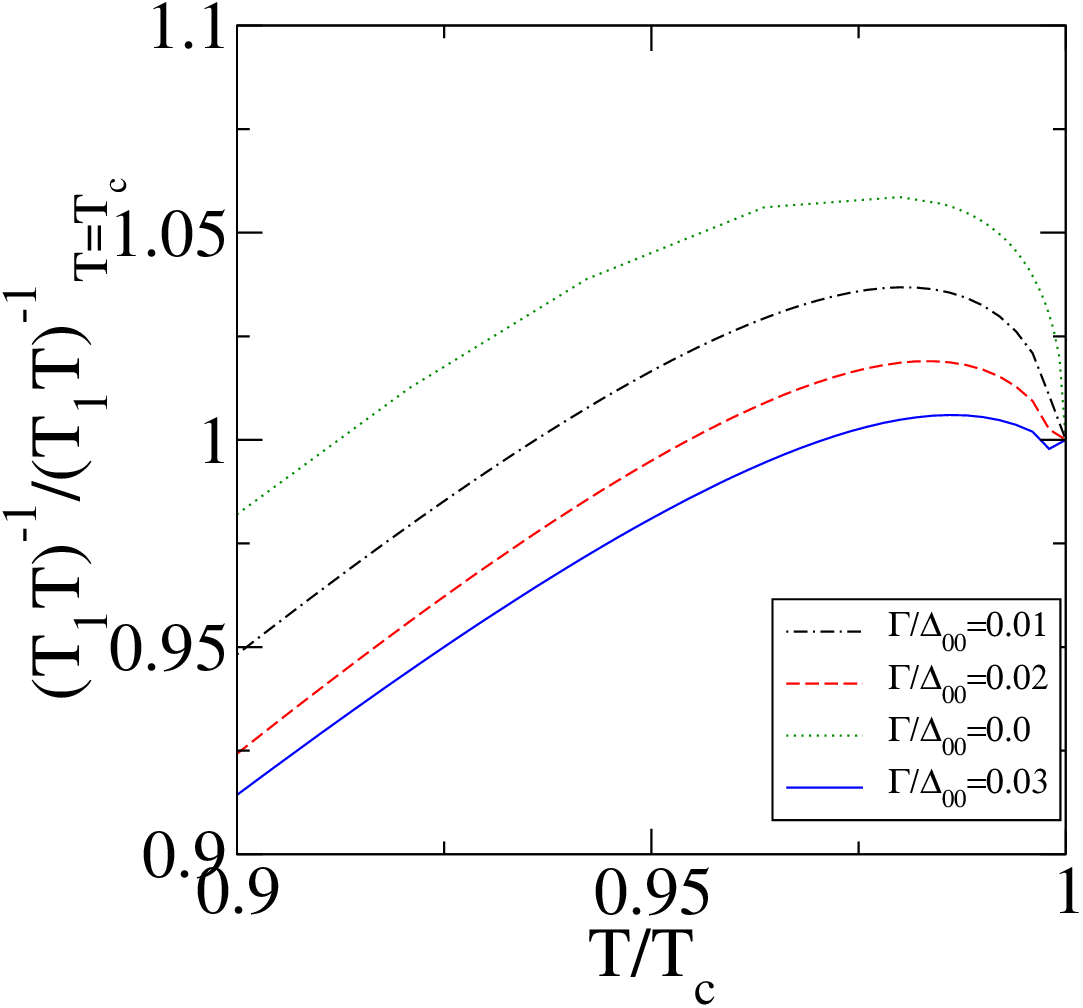}
\caption{The numerically computed T$_{1}^{-1}$ for no impurity scattering,
and for $\Gamma/\Delta_{00}$ = 0.01, 0.02 and 0.03 are shown.}
\end{figure}
concentration required to destroy superconductivity \cite{AG}, and would
result in a depression of T$_{c}$ of this order.  As is
clear from the plot, the height of the peak is greatly diminished even 
by the rather low impurity scattering \cite{hirschfeld} rates modeled here.

Given that materials in the unitary limit typically have superconductivity
destroyed by an impurity concentration on the order of a few percent 
\cite{maki_redbook}, the foregoing analysis indicates that an impurity 
concentration of just 0.25 percent is sufficient to largely destroy this
peak.  Such a concentration is well within the range of observation \cite{graf_UPt}. 

In order to observe this peak samples of the highest possible
quality are clearly essential, with impurity concentration less than
0.1 percent.  It would also be advantageous to perform
low-temperature specific heat measurements on the same samples as this
would allow accurate assessment of the prediction of a finite relaxation
rate at T=0, via a measurement of the residual density of states.

An additional effect complicating the observance of this peak is the frequent
occurrence of magnetism in the heavy-fermion and high-T$_{c}$ cuprate
materials upon which most of the measurements have been performed.
The combination of magnetism and the extreme sensitivity of this peak to the
presence of impurities in the unitary limit make its observation difficult in
the heavy-fermion and cuprate superconductors.  
However, the recently discovered non-centrosymmetric superconductor Li$_{2}$Pt$_{3}$B \cite{badica} shows no signs of magnetism or strong electron correlation \cite{yuan},
and appears to be unconventional on at least one band, based upon magnetic penetration depth data \cite{yuan}.  This material may therefore be an ideal material in which to search for this small T$_{1}^{-1}$ peak.
Another possibility for experiment is the class of organic 
superconductors, which may not necessarily have the 
sensitivity to impurities characteristic of the heavy-fermion and cuprate materials. 
Indeed, the organic superconductor (TMTSF)$_{2}$PF$_{6}$ has already shown
a small peak below T$_{c}$ \cite{lee}.

To summarize, here we have demonstrated that any unconventional superconductor
obeying BCS pure-case theory should show a small coherence peak in the nuclear
spin lattice relaxation rate T$_{1}^{-1}$ just below T$_{c}$.  It is likely due
to magnetic and impurity effects that this peak has not generally been observed in unconventional superconductors.

{\bf Acknowledgment}

We are indebted to K. Maki for useful suggestions.

\end{document}